\documentclass[useAMS,usenatbib]{mn2e}
\usepackage{epsfig}
\def\lesssim{\mathrel{\hbox{\rlap{\hbox{\lower4pt\hbox{$\sim$}}}\hbox{$<$}}}}
\def\gtrsim{\mathrel{\hbox{\rlap{\hbox{\lower4pt\hbox{$\sim$}}}\hbox{$>$}}}}
\def\apj{ApJ}

\def\aj{AJ}
\def\aap{A\&\hskip-1pt A}

\title[Structure of  open clusters]{Deep and Wide Photometry of Two Open
Clusters NGC 1245 and NGC 2506: Dynamical Evolution and Halo}
\author[Lee, Kang \& Ann]{S. H. Lee,$^1$\thanks{E-mail:
ngc2420@hanmail.net} Y.-W. Kang$^1$ and H. B. Ann$^2$\thanks{Author to whom any
correspondence should be addressed. E-mail:hbann@pusan.ac.kr}\\
$^1$Korea Astronomy and Space Science Institute, Daejeon 305-348, Korea\\
$^2$Department Earth Science Education, Pusan National University, Busan 609-735, Korea}

\begin{document}

\date{Accepted 2013 April 5. Received 2013 April 4; in original form 2012 July 19
}
\pagerange{\pageref{firstpage}--\pageref{lastpage}}
\pubyear{2013}

\maketitle
\label{firstpage}

\begin{abstract}

We studied the structure of two old open clusters, NGC 1245 and NGC 2506, from 
a wide and deep $VI$ photometry data acquired using the CFH12K CCD camera 
at CFHT. We devised a new method for assigning cluster membership probability 
to individual stars using both spatial positions and positions in the 
colour-magnitude diagram. From analyses of the luminosity functions at several
cluster-centric radii and the radial surface density profiles derived from 
stars with different luminosity ranges, we found that the two clusters are
dynamically relaxed to drive significant mass segregation and evaporation
of some fraction of low-mass stars. 
There seems to be a signature of tidal tail in NGC 1245 but the signal is too
low to be confirmed.

\end{abstract}

\begin{keywords}
galaxies: star clusters -- open clusters: general -- method: observation --techniques: photometric
\end{keywords}

\section{Introduction}

Open clusters are useful examples for studying the dynamical evolution of 
stellar systems and the chemical evolution of the Galactic disc,
because the stars in an open cluster are assumed to have the same chemical
and dynamical properties owing to of their shared origin.
Because open clusters orbit about the centre of the Galaxy near the Galactic 
plane, they are frequently disturbed by both giant molecular clouds and
the tidal forces originating from the Galactic disc. In particular,
old open clusters can be tracers of the structural and evolutionary 
history of the Galactic disc \citep{fri95}.

The structure of the old open clusters can be changed by dynamical
evolution driven by internal and external tidal forces.
Energy equipartition due to encounters among member stars \citep{spi40} leads 
to mass segregation, which is characterized by the concentration of high-mass
stars in the cluster's centre, and an isotropic distribution of low-mass
stars rambling throughout the cluster. Mass segregation results in the steeper
slope of the mass function in the outer region of the cluster \citep{ann02}.
Some fraction of low-mass stars evaporates from the cluster to create the halo
surrounding the cluster \citep{egg93}.
The evaporation of low-mass stars also increases the size of the cluster 
because of the reduction of the binding energy.

\begin{table*}
 \centering
 \begin{minipage}{145mm}
 \raggedright
  \caption{Physical parameters of NGC 1245 and NGC 2506 taken from Paper I}
   \begin{tabular}{@{}ccccccccc@{}} \hline
  Cluster   &$\alpha$ &  $\delta$  &       $l$   &   $b$     & $E(B-V)$  & $(V - M_v )_0 $& age(Gyr)&  [Fe/H]   \\
    name   & J2000.0   &   J2000.0    &   (deg.)   &  (deg.)  &                 &                        &                &                \\ \hline\hline
NGC 1245 &  03 14 47 &  +47 13 53 &  $146.65$ & $-8.93$ &     0.24     &      12.25           &    1.08     &     -0.08       \\
NGC 2506 &  08 00 02 &  -10 46 15 &  $230.56$ & $+9.94$ &     0.035    &      12.47          &    2.31      &   -0.24        \\
   \hline
  \end{tabular}
  
Notes. -- Units of right ascension are hours, minutes, and seconds; units of declination are degrees, arcminutes, and arcseconds.

 \end{minipage}
\end{table*}

Although internal interactions among member stars make the distribution of stars
more spherical,
perturbation by Galactic tidal forces results in an elongated shape.
The elongated morphology of open clusters has been reported for
the Hyades \citep{oor79},
Pleiades \citep{rab98a}, NGC 2287, and NGC 2548 \citep{ber01}.
\citet{che04} also reported the elongated structure of 31 open clusters,
which includes young star clusters that are a few million years old.
\citet{sha06} reported the elongated corona of 6 open clusters among 9 open
clusters. They could not find a correlation between the age of a cluster and its core shape.

One possible origin of the elongated morphology of open clusters is a
stretching of the stellar distribution along the cluster orbit \citep{chu10}.
\citet{dav10} represented the halo morphology of M67 is elongated shape using SDSS data, 
which is roughly aligned with the proper motion of the cluster.
A relevant structure is the tidal tail caused by the Galactic tidal field \citep{chu06}.
Because the evaporation of low-mass stars is likely to be isotropic,
if there is no external force,
the elongated shape of the halo of open clusters seems to be caused
by the same mechanism that creates the tidal tails.
Thus, it seems likely that a detailed analysis of halo structure provides 
important information about the dynamical history of open clusters.
However, analysis of the spatial structure of open clusters is not easy,
because it is extremely difficult to assign cluster membership
without having kinematic information.

The membership of cluster stars is critical for any investigation of the 
dynamical structure of clusters.
Earlier studies of dynamical evolution and halo structure were carried out 
for nearby clusters whose member stars were well defined by proper motion
studies \citep{mer90, per98, rab98a, rab98b}.
Since the kinematic data for membership determination is difficult to be
attained for most of open clusters, especially for the faint stars, 
the colour-magnitude diagram (CMD) of clusters and surrounding field regions 
are used to correct 
the field star contamination in the cluster luminosity 
function  \citep{sun99}. There have been some attempts to determine
membership probability of individual stars statistically using spatial 
densities \citep{che04} or stellar densities in CMD \citep{jan11}. 
Here, we introduce a new method for assigning cluster membership probability
based on locations in the CMD and projected spatial positions in the sky. 

The aim of the present study is to understand the structures and dynamical 
properties of two old open clusters, NGC 1245 and NGC 2506, for
which $VI$ CCD photometry data has been reported by \citet{lee12} (Paper I).
Because the photometry was deep and wide enough to analyze the surface 
density distribution of low-mass stars around the tidal radii of NGC 1245
and NGC 2506, we expect to be able to describe the halo structures of the two
clusters. We also expect to see the effects of dynamical evolution on the
structure of open clusters when analyzing the radial dependence of LFs.

In \S2, we describe the observational data and the cluster parameters.
We describe membership probability in \S3.
The properties of dynamical evolution and halo are analyzed in \S 4.
A brief summary and conclusions are given in the final section.

\section{OBSERVATIONAL DATA} 

The data used in the present study are from the deep and wide field
$VI$ CCD photometry observations we conducted in a previous study (Paper I).
A detailed description of the observations and data reduction are given in Paper I; 
however, we briefly describe the photometry here.
We obtained the data using a CFH12K mosaic CCD camera mounted at the prime
focus of the CFHT.
The CFH12K CCD camera has the field of view $42^{\prime}\times28^{\prime}$ 
with a pixel size of $0.^{\prime\prime}206$.
We observed 6 regions in NGC 1245 and 3 regions in NGC 2506 (all of which 
were centred on the cluster centres),
and which covered $84^{\prime} \times 82^{\prime}$ and $42^{\prime}\times 81^{\prime}$ 
for NGC 1245 and NGC 2506, respectively.

We followed the standard procedures for CCD reduction using IRAF and
carried out PSF photometry using DAOPHOT \citep{ste87}.
The seeing size, defined as the FWHM of the stellar images, varied between
$1.^{\prime}1$ and $1.^{\prime}6$ from region to region.
For calibrating our photometry, we observed the SA 98 region \citep{lan92} on the same night.
The limiting magnitude of the photometry is V $\sim$ 23.
The cluster parameters are listed in Table 1; they were derived using the CMDs as described in Paper I.

\begin{figure*}
 \vspace{-58mm}
 \hspace{-6mm}
 \epsfig{figure=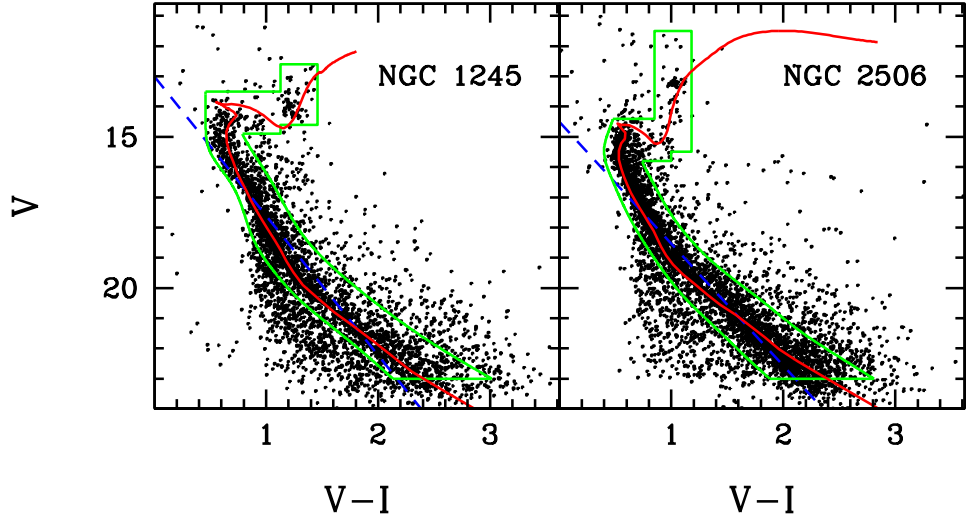, height=0.92\textwidth, width=0.92\textwidth}
 \vspace{-4mm}
 \caption{Color-magnitude diagrams of NGC 1245 and NGC 2506 with the stars 
in the central region ($ R < 10^{\prime}$).
Green solid lines in the panels indicate the photometric membership 
boundaries of the cluster stars. The blue dashed line indicates the rough
positions of the main sequence to divide the non-member stars into those
above the main sequence and those below the main sequence.
}
 \label{CMD_membership}
\end{figure*}
\begin{figure*}
 \vspace{-18mm}
 \epsfig{figure=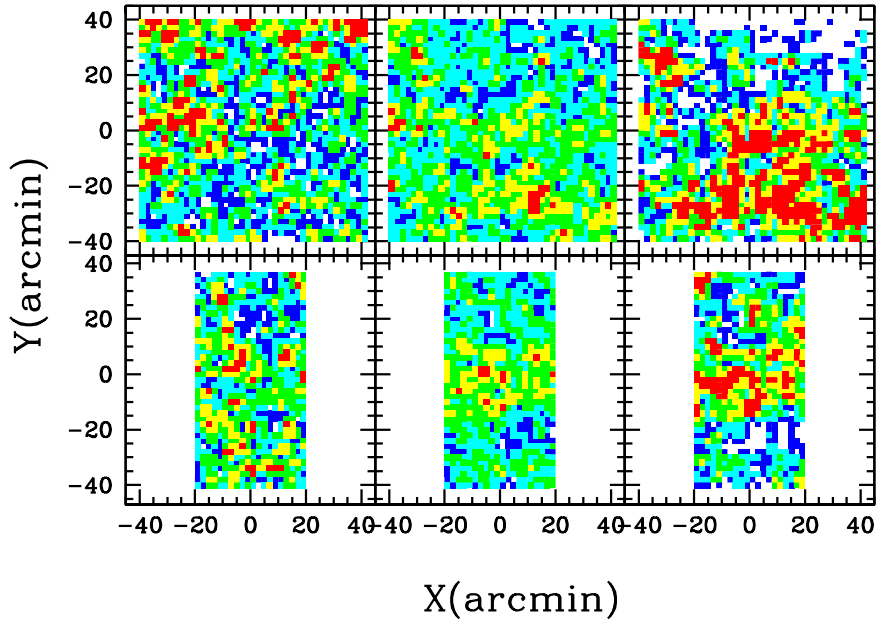, height=0.8\textwidth, width=0.8\textwidth}
 \vspace{-20mm}
 \caption{The normalized spatial density distribution of field stars which 
are distributed in the outer region of the membership boundary in Fig. 1. 
Upper pannels are for NGC 1245 and lower pannels are for NGC 2506. 
The left pannels and right pannels show the distributions of the field stars 
located above and below the blue dashed lines in Fig. 1, respectively.
The middle pannels represent the density distribution of all field stars which 
are located outside the membership boundaries in Fig 1. 
The colors represent the density increment of $\Delta g(x,y) = 0.14$ from
the lowest level as blue color and the highest level as red color
 with three intermediate
levels, cyan, green and yellow among which green color represents the density
level of $g(x,y)$=1. 
}
 \vspace{0mm}
 \label{chart_P}
\end{figure*}


\section{MEMBERSHIP PROBABILITY}

\begin{figure}
 \vspace{-4mm}
 \epsfig{figure=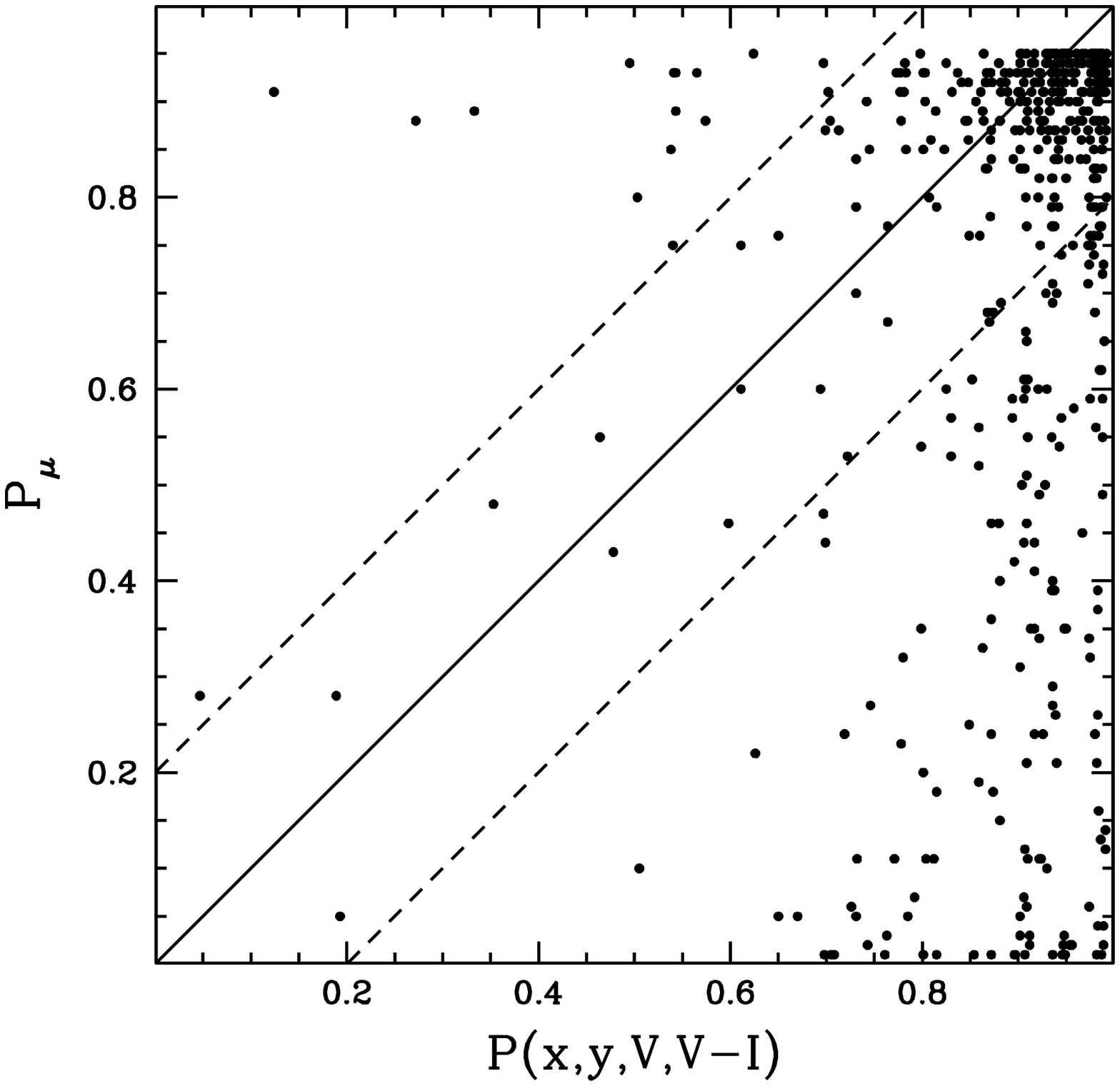, height=0.5\textwidth,width=0.5\textwidth}
 \vspace{-2mm}
 \caption{
Correlation between proper motion membership \citep{chi81} and
photometric statistics for stars in NGC 2506. Stars between dotted lines have
membership differences smaller than 20\%.
}
 \vspace{0mm}
 \label{rad_nod}
\end{figure}

Traditionally, proper motion study is the most reliable method for 
determining the membership criteria of cluster stars. However,
this method requires two observations separated by a long time interval.
Therefore, most previous studies utilized membership criteria 
based on the locations of cluster stars in the CMDs \citep{sun99}.
This approach is very simple and effective when the main sequences do not 
overlap with wedge patterns of field stars in the CMD.

The main sequence and giant stars of a cluster are distributed 
along the cluster isochrone in the CMDs,
whereas non-member stars are primarily distributed in a wedge shape 
with a random distribution outside the wedge pattern.
There is also a clear difference in the spatial distributions of 
the cluster member stars and those of non-member stars.
The cluster stars are likely to be centrally concentrated, whereas field stars
are uniformly distributed, which leads to an outward decrease of the ratio of
the number of member stars to the number of field stars.
Thus, it seems plausible that we can differentiate the cluster members 
from the field stars using CMDs and the spatial distributions 
of stars in the cluster region.

According to the stellar evolution theory, cluster members are distributed 
along the main sequence and giant branches in the CMD. 
To determine the boundaries of the two clusters within which the member stars 
are supposed to be located in the CMD, we derived the density distributions 
across the main sequence at each magnitude bin using stars in the central 
region, $ R < 10^{\prime}$ and fitted Gaussian functions to the density
cross-sections along the main sequence.
We smoothly connected the locations correspond to the 3 $\sigma$ density levels 
to determine the photometric membership boundary in the CMDs (Fig. 1). 
For the giant stars we defined the boundaries based on the visual inspection 
of the giant stars in the CMD. 
We neglected the blue stragglers in defining the membership boundaries. 
The magnitude limits are $V\approx23$ for the clusters.

We define the membership probability of a star $i$, $P_{i}$, as a function of
the coordinates in the four-dimensional (4-D) space that consists of 
two spatial coordinates in the sky and two photometric parameters 
constituting the CMD, namely $x, y, V, V-I$.
Thus, we consider a 4-D membership probability as follows. 
We counted the number of stars in a unit volume of the 4-D space centered 
on a point $(x, y, V, V-I)$ and then corrected for the contribution by field 
stars. We derived the field star distribution in CMD, i.e., $N_{f}(V, V-I)$,
to correct for field star contamination using the stars in the 
outermost regions $R > 40^{\prime}$ for NGC 1245
and $R > 35^{\prime}$ for NGC 2506. 
If we assume that the distribution of field stars is homogeneous in the 
cluster region,
the distribution function of cluster member stars in the 4-D space,
$N_{c}(x,y,V,V-I)$, is determined from the following equation:
\begin{equation}
N_{c}(x,y,V,V-I)=N(x,y,V,V-I)-N_{f}(V,V-I)
\end{equation}
where $N(x,y,V,V-I)$ is the distribution function of stars in the
4-D space, including both member and non-member stars.

Because field stars are mostly disc stars which are located in the 
Galactic plane where considerable density variation is expected,
it is better to take into account the density variation of field stars.
To investigate the variation of the spatial distribution of non-member stars, 
we derived the spatial distribution of field stars, using stars outside the
cluster boundary, by calculating the 
normalized field star density distribution $g(x,y)$ defined as
\begin{equation}
g(x,y) = n(x,y)/<n(x,y)>
\end{equation}
where $n(x,y)$ is the spatial number density of stars outside the
photometric membership boundaries in the CMD and $<n(x,y)>$ is 
the average value of n(x,y). 
Thus, the mean value of $g(x,y)$ is equal to 1. 
We derived $n(x,y)$ by counting the number of stars in the smoothing box
of $4^{\prime} \times 4^{\prime}$, with sampling interval 
of $2^{\prime} \times 2^{\prime}$. These sampling size and sampling interval
were used through out the paper for the derivation of the surface number 
densities such as N (x,y,V,V-I) and g(x,y). The sampling size and sampling
interval in the CMD are $\Delta V=1.04, 0.52$ and $\Delta (V-I)=0.269, 0.134$.
We used a Gaussian smoothing algorithm in 4-D space to derive the surface 
number densities.

\begin{figure}
 \vspace{-2mm}
 \hspace{-10mm}
 \epsfig{figure=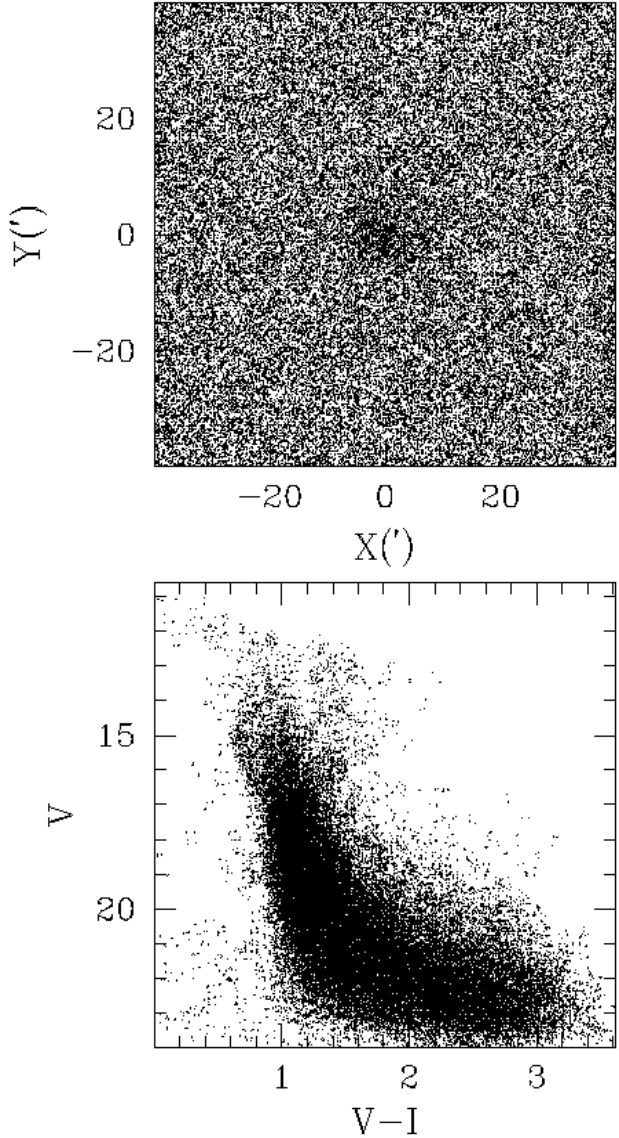, height=1.0\textwidth,width=0.55\textwidth}
 \vspace{-6mm}
 \caption{
Spatial distribution and CMD of an artificial cluster generated
by Monte-Carlo simulation for NGC 1245.
}
 \vspace{-2mm}
 \label{rad_nod}
\end{figure}

Fig. 2 shows the spatial distribution of non-member stars, represented by
$g(x,y)$. The distributions of field stars below and above the main sequence
are plotted in the right panels and left panels, respectively. 
The middle panels show the spatial distribution of all the field stars in the 
CMDs.
The upper panels are for NGC 1245 and the lower panels are for NGC 2506.   
There are some differences in the spatial distributions of the stars 
in the right panels and left panels. In particular, the stars in the upper
right panel, i.e, field stars below the main sequence of NGC 1245 show higher
density in the middle south region of the cluster 
while those above the main sequence of NGC 1245 
show higher density in the north-east part of the cluster.   
We suppose that this difference is caused by the gradient 
in the density distribution of field stars toward NGC 1245.  
The stars above the main sequence are likely to be foreground stars 
while those below the main sequence are likely to be background stars 
because the stars above the main sequence are brighter than those 
below the main sequence if they have similar colors.

Using $g(x,y)$, we rewrote the 4-D distribution function of cluster 
member stars $N_{c}(x,y,V,V-I)$  as \\\\
$N_{c}(x,y,V,V-I) = N(x,y,V,V-I)$
\begin{equation}
\qquad \qquad \qquad \qquad \quad   -N_{f}(V,V-I)g(x,y)   
\end{equation}
Then, the membership probability, $P_{i}$, of a star $i$ is defined as
\begin{equation}
P_{i} = {{N_{c}(x,y,V,V-I)} \over N(x,y,V,V-I)}
\end{equation}
We derived the membership probability of all stars in the observed regions of
the two clusters, however, we assigned $P_{i}=0$ to the stars outside 
the photometric membership boundary in the CMDs (Fig. 1).

To check the validity of the derived values of $P_{i}$, 
we compared the membership probabilities of stars in NGC 2506
derived from the present method using $P(x,y,V,V-I)$ and $P_{\mu}$ of
\citep{chi81} in Fig. 3. 
\citet{chi81} derived the membership probabilities $P_{\mu}$ of 724 stars in
NGC 2506 using two photographic plates with an epoch difference of 56.9 years.
Their sky coverage is only $5.^{\prime}3$ in radius, with a magnitude limit of
$V \approx 16$.
As can be seen in Fig. 3, the majority of stars with a high membership
probability ($P_{\mu} > 0.7$) also have a high $P(x,y,V,V-I)$ value. 
However, a significant fraction of high $P(x,y,V,V-I)$ stars
have a $P_{\mu}$ less than 0.7. This is because the proper 
motion study was confined to the stars in the central regions of the cluster,
where high values of $P(x,y,V,V-I)$ are assigned to stars if their CMD 
positions are within the photometric membership boundary.  
Thus, the present method of determining membership probabilities is likely to 
overestimate the membership probabilities for stars in the central region.

In order to see the robustness of segregating the cluster member
stars from the field stars,
we created artificial clusters using the radial density profile and 
the mass function of NGC 1245 via the Monte-Carlo technique. 
The spatial distribution of each star was determined 
by randomly generated radial positions assuming the King profile 
whereas the positions in the CMD were determined from the luminosity function 
assuming asymmetric Gaussian distribution for the colors at a given luminosity. 
We did not assume binary sequence explicitly but we took into account the 
binary effect 
by assuming three times larger $\sigma$ for the Gaussian distribution 
for the color distributions in the right side of the main sequence. 
We also generated the artificial field stars by assuming homogeneous
spatial distribution and the CMD positions of the field stars which are
located at $R > 40^{\prime}$ from the cluster centre. 
We derived the CMD position of each star by assigning magnitude and color 
randomly around each field star position in the CMD with $\Delta V=0.125$ 
and $\Delta (B-V)=0.025$.
We constrained the number of field stars by comparing the area where the
field stars were selected and the total observed field of NGC 1245.
In Fig. 5, we presented the spatial distributions and CMDs of an artificial
cluster including artificial field stars.

We ran 100 simulations to see that the 4-D membership probability 
technique works correctly. 
Fig. 5 shows the fractional frequency distributions of the cluster members and 
non-members as a function of the 4-D membership probability $P$. 
Most of the field stars have $P$ less than $\sim0.2$ with $\sim70\%$ of 
field stars have $P=0$ whereas $\sim50\%$ of the cluster members have $P$ 
larger than 0.2. 
As can be seen in Fig. 5, if we randomly select a star with $P=0.15$, 
the probability that this star is a field star is $\sim50\%$. 
But, if we randomly select a star with $P=0$, the field star 
probability is $\sim70\%$ whereas if we select a star with $p>0.4$, that is
less than $\sim10\%$. The reason for the relatively high number of member 
stars that have $P=0$ is the number of member stars is only $4\%$ of the
field stars. This makes $P=0$ for member stars which are located in the 
regions where the surface density is greatly dominated by field stars. 
A predominance of field stars makes $N_{c}\approx0$ and consequently 
$P$ becomes zero (see eqs (3) and (4)). Thus, the member stars that 
have $P=0$ are mostly located in the region where field stars are distributed
with a wedge pattern in the CMD.

\begin{figure}
 \vspace{-3mm}
 \hspace{-6mm}
 \epsfig{figure=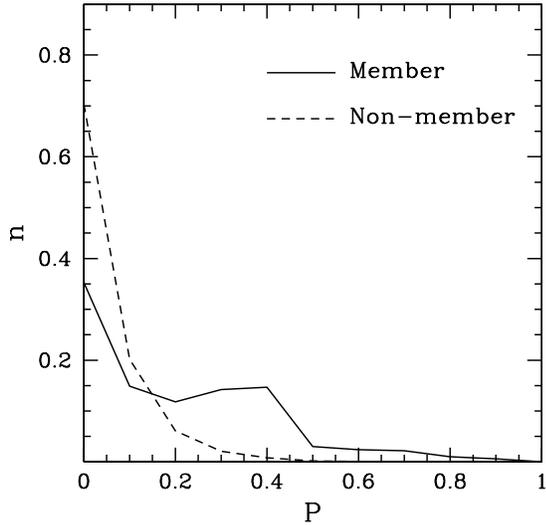, height=0.52\textwidth,width=0.52\textwidth}
 \vspace{-6mm}
 \caption{
Fractional frequency distribution of membership probability of the member
and non-member stars. Solid and dotted lines represent member stars and
non-member stars, respectively.
}
 \vspace{3mm}
 \label{rad_nod}
\end{figure}

\section{STRUCTURE AND DYNAMICAL EVOLUTION OF OPEN CLUSTERS}
\subsection{Radial Surface Density Profiles}

Similar to \citet{che04}, we defined an effective surface number density 
by summing up the membership probabilities, $P_{i}$, inside the observed area $\Delta S$,
\begin{equation}
f={1\over{\Delta S}}{\Sigma P_{i} \over \Lambda}
\end{equation}
where $\Lambda$ is the correction factor for incomplete photometry (Paper I).
Because the position of maximum value of $f$ ($f_{max}$) depends upon 
the size of $\Delta S$,
we first determined the position of $f_{max}$ by calculating $f$ for all 
the pixel positions $(x, y)$ with $\Delta S=1^{\prime} \times 1^{\prime}$.
Then we determined the position of the cluster centre by averaging 
the coordinates of the pixels where $f$ is larger than $0.9f_{max}$.
We derived the radial surface number density $f(R)$ by calculating $f$ for 
concentric rings at a radial distance $R$ with radial interval of
$dR=(1.1)^{n} dR_{0}$ where n represents the nth ring and
$dR_{0}=1.^{\prime}0$ is the radius of the central region. Although $f$
is considered to be the surface number density corrected for field stars,
there are some residuals, $f_{bg}$ due to field stars which have positive 
$P_{i}$ owing to their CMD positions lying within the cluster membership 
boundaries. We calculated $f_{bg}$ at $R > 40^{\prime}$ for NGC 1245 and 
and $R > 35^{\prime}$ for NGC 2506, respectively. Because $P_{i}$ depends on
the smoothing box and sampling resolution, we optimized them to minimize
$f_{bg}$. The adopted values of $f_{bg}$ are $0.20\pm0.62$ for NGC 1245 and
$0.25\pm0.63$ for NGC 2506. We subtracted $f_{bg}$ from $f$ before fitting to
the King profiles \citep{kin62}.

Fig. 6 and  Fig. 7 show the logarithmic radial surface number density profiles 
of NGC 1245 and NGC 2506 with the best-fit King profiles 
which can be expressed as    
\begin{equation}
f(R)={f_{o} \over {1+(R/R_{c})^{2}}} 
\end{equation}
where $f_{o}$ is the central surface number density, $R_{c}$ is core radius.
We derived $f_{o}$ and $R_{c}$ by least-squares fitting technique.
The radius for the cluster extent was taken to be $R_{cl}$, where the 
radial surface number density profiles becomes constant.   
We derived $R_{cl} = 16^\prime$ and $19^\prime$ for NGC 1245 and NGC 2506, 
respectively (paper I).   

\begin{figure}
 \vspace{4mm}
 \hspace{-42mm}
 \epsfig{figure=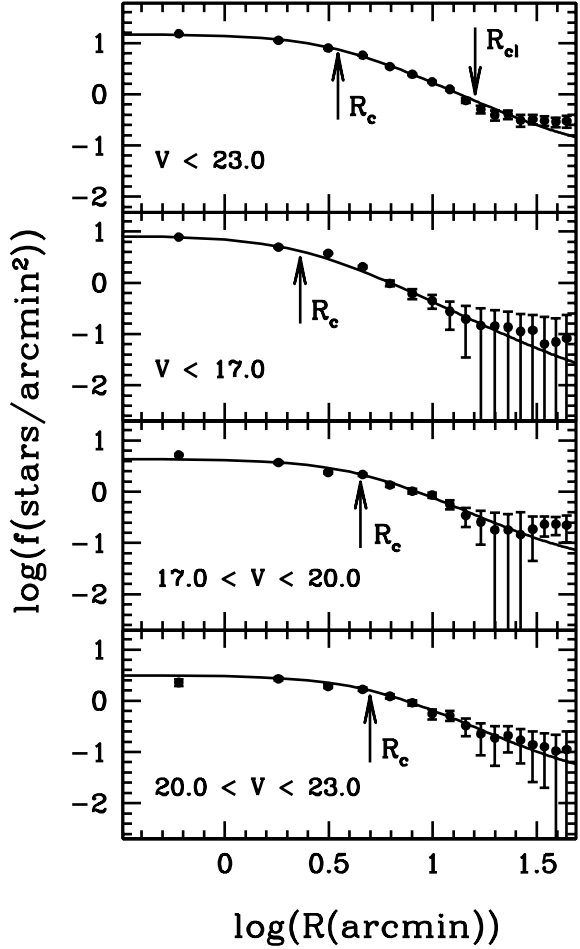, height=0.95\textwidth, width=0.95\textwidth}
 \vspace{-9mm}
 \caption{
King model fitting to the radial surface density profiles for NGC 1245.
Solid curve represents best fit \citet{kin62} to the observed data.
The error bars are Poisson errors.
}
 \vspace{3mm}
 \label{rad_halo}
\end{figure}

\begin{figure}
 \vspace{4mm}
 \hspace{-42mm}
 \epsfig{figure=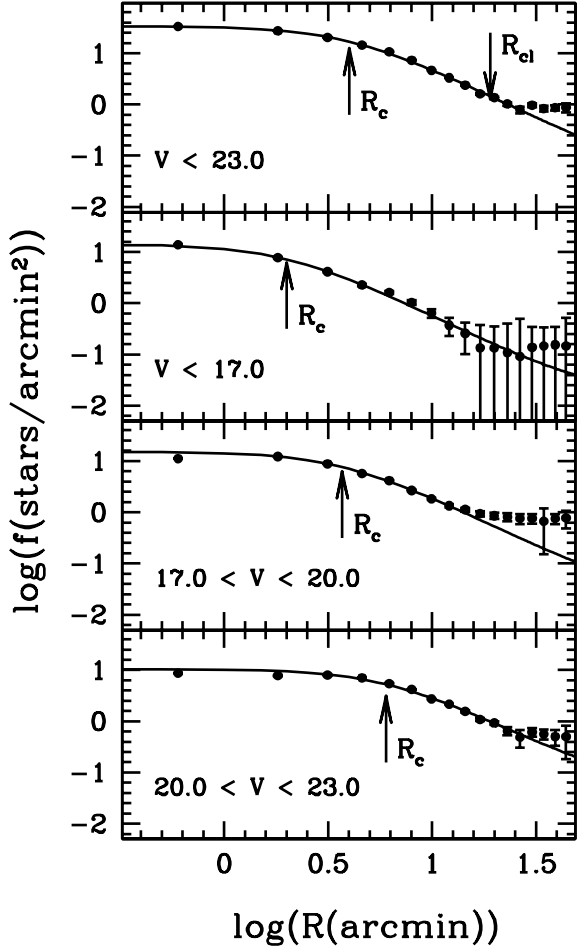, height=0.95\textwidth, width=0.95\textwidth}
 \vspace{-9mm}
 \caption{
King model fitting to the radial surface density profiles for NGC 2506.
Solid curve represents best fit \citet{kin62} to the observed data.
The error bars are Poisson errors.
}
 \vspace{3mm}
 \label{rad_halo}
\end{figure}

Since the effective radial surface number density near $R_{cl}$ is 
comparable to $f_{bg}$, errors of $f_{bg}$ affect the derived $f_{o}$ 
and $R_{c}$ significantly. For NGC 1245, $1\sigma$ variation of $f_{bg}$ 
gave rise to $\sim30\%$ and $\sim50\%$ variations in $f_{o}$ and $R_{c}$ 
while they were reduced to $\sim10\%$ to both $f_{o}$ and $R_{c}$ for NGC 2506.

We examined the dependence of $R_{c}$ on the luminosity of stars 
because the spatial distribution of stars depends on the stellar mass 
if the clusters under consideration are dynamically relaxed. 
We plotted the effective radial surface number density profiles 
along with best-fitted King profiles to show this dependence in Fig. 6 
and Fig. 7. 
As expected, the effective radial surface number density profiles of 
bright stars have smaller core radius than those of faint stars (see Table 2).
This is the very clear evidence of the mass segregation due to 
dynamical evolution of the clusters.

One thing worth to mention is the departure of 
the observed surface number density from the King profile 
at $R > 15^{\prime}$, which is more pronounced in NGC 1245. 
Although the difference between the observation and the King profile is
within the Poisson errors, there is a possibility that they are caused by 
some real features such as tidal tails, especially in NGC 1245. 
We can see the morphology of these features in Fig. 9 where we found 
that there are over-dense regions at $R\sim30^{\prime}$ 
in the north-west of NGC 1245.
However, the surface density of the over-dense region is about 
$3\sigma$ of the $f_{bg}$, the morphology of the over-dense regions are
much affected by the errors in $f_{bg}$.

\begin{table}
\centering
\vspace{10pt}
 \begin{minipage}{145mm}
 \raggedright
  \caption{King model fitting parameters for NGC 1245 and NGC 2506}
  \hspace{3mm}
   \begin{tabular}{@{}cccc@{}} \hline

  cluster  & Magnitude range   & $f_{o}$ &  $R_{c}$  \\
   \hline  
  NGC 1245 &  $     V < 17$    &   8.3   &   2.3    \\
           &  $17 < V < 20$    &   4.3   &   4.5    \\
           &  $20 < V < 23$    &   3.1   &   5.0    \\
           &  $     V < 23$    &  14.7   &   3.6    \\
  NGC 2506 &  $     V < 17$    &  14.3   &   2.0    \\
           &  $17 < V < 20$    &  15.2   &   3.7    \\
           &  $20 < V < 23$    &  10.4   &   6.0    \\
           &  $     V < 23$    &  33.5   &   4.0    \\
   \hline
  \end{tabular}
 \end{minipage}
\end{table}

\subsection{Radial LFs}

We derived the luminosity function (LF) of the clusters using 
the membership probability defined in equation 4. 
Because the number of stars at each magnitude interval, 
corrected for the field star contamination,
is equal to the effective number of stars $N_{eff}$, it can be expressed as
\begin{equation}
N_{eff}(m)={\Sigma P_{i}(m) \over \Lambda(m)}
\end{equation}
where $P_{i}(m)$ and $\Lambda(m)$ are the membership probability and 
incompleteness correction factor,
respectively, for stars with a magnitude interval $(m-0.5 \sim m+0.5)$.
Then, the LFs at a radial distance $R$ from the cluster centre can be 
determined based on $N_{eff}(m)$.

We considered three regions, $R < R_{c}$, $R < R_{1/2}$,
and $R_{1/2} < R < R_{cl}$, to derive Radial dependence of LFs. 
The half number radius $R_{1/2}$ is the radius where the number of 
stars becomes half of the total number of stars within $R_{cl}$.
They are  $R_{1/2} = 7.^{\prime}1$ and $7.^{\prime}8$ 
for NGC 1245 and  NGC 2506, respectively. 

\begin{figure}
 \hspace{-7mm}
 \vspace{-3mm}
 \epsfig{figure=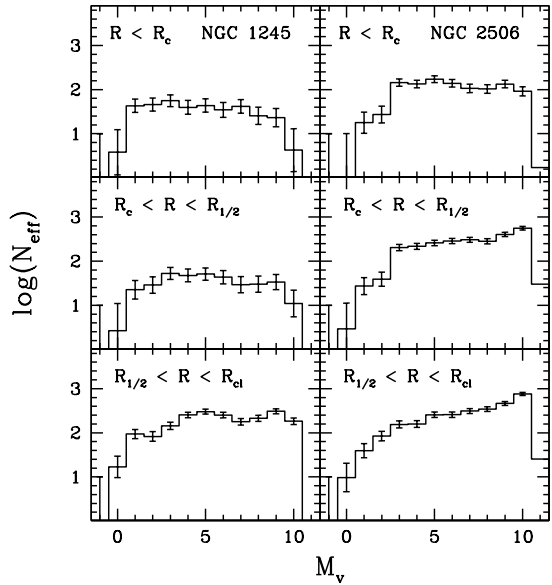, height=0.52\textwidth, width=0.52\textwidth}
 \vspace{-3mm}
 \caption{
 Radial dependence of the luminosity functions of NGC 1245 and NGC 2506.
Clusters are divided into three regions, $R<R_{c}$, $R_{c}<R<R_{1/2}$, and
$R_{1/2}<R<R_{cl}$, from the left to the right.
The error bars are Poisson errors.
}
 \label{rad_LF}
\end{figure}

Fig. 8 shows the radial LFs derived for the two open clusters, NGC 1245 and NGC 2506.
The general trend of the LFs of the two open clusters under consideration is 
a gradual change of the peak positions; the LFs peak closes to 
the bright end of the inner region ($R < R_{c}$), 
whereas the peak closes to the faint end in the outer region ($R_{1/2} < R < R_{cl}$).
This means that fainter stars are likely to be found in the outer regions
of the clusters, i.e., there is a clear evidence of mass segregation
in the two clusters, NGC 1245 and NGC 2506.

Similar signatures of mass segregation were reported  for M11 \citep{mat84},
Pleiades and Praesepe \citep{rab98a, rab98b}, NGC 2099 \citep{kal01b}
and NGC 6819 \citep{kal01a, kan02}).
The reason for the observed mass segregation seems to be the dynamical evolution
that leads to the energy equilibrium  because these open clusters are old 
enough to be relaxed.

\begin{figure}
 \vspace{0mm}
 \hspace{-4mm}
 \epsfig{figure=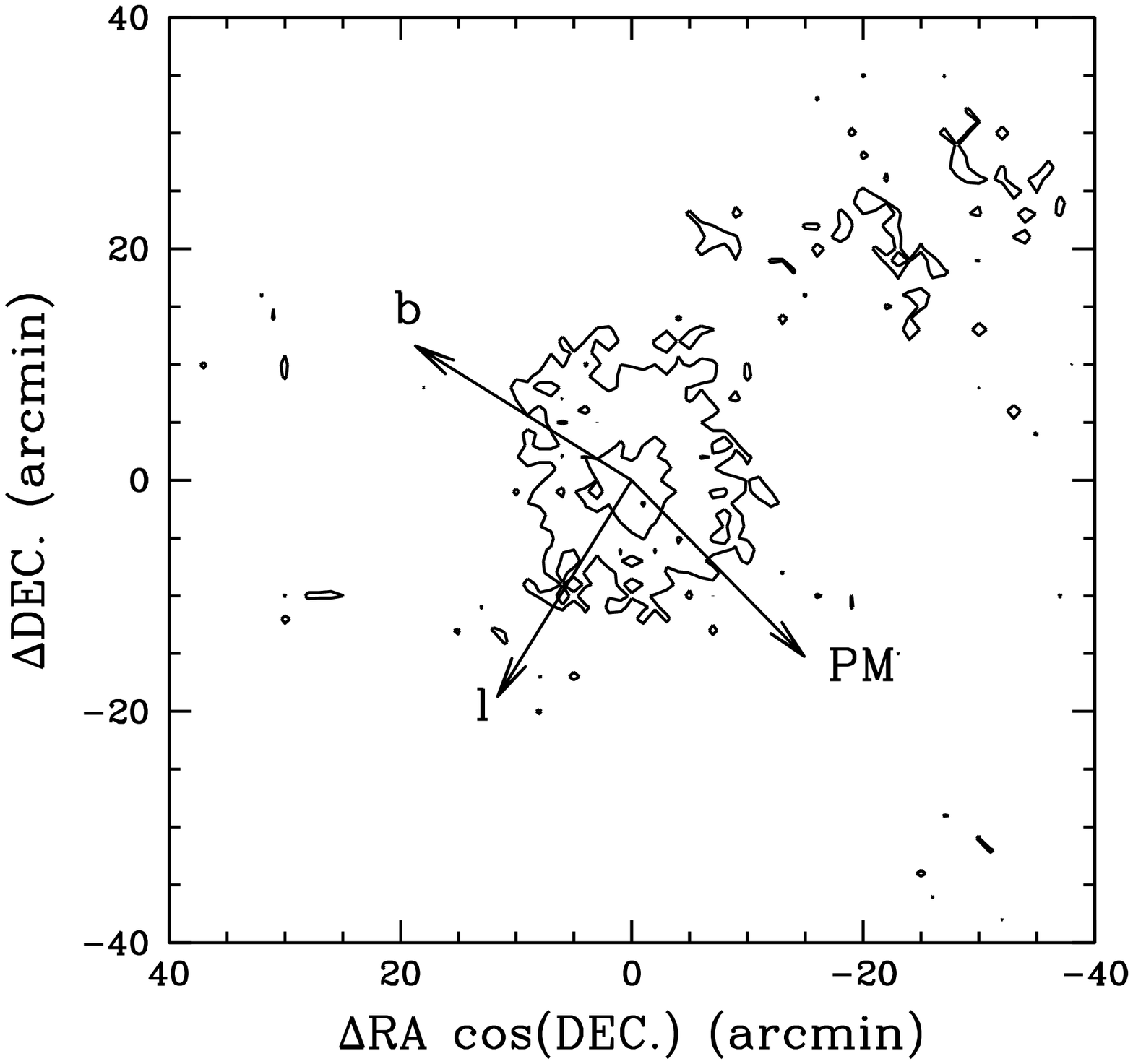, height=0.52\textwidth,width=0.52\textwidth}
 \vspace{-6mm}
 \caption{
Log scale density contour map for NGC 1245. 
The inner contour represents $f_{o}/2$ and the outer contour indicates 
$3\sigma$ level of background density fluctuation.
Each arrows represent the Galactic coordinate (l, b) and the proper 
motion vector (PM).
}
 \label{density_contour}
\end{figure}

\begin{figure}
 \vspace{0mm}
 \hspace{-4mm}
 \epsfig{figure=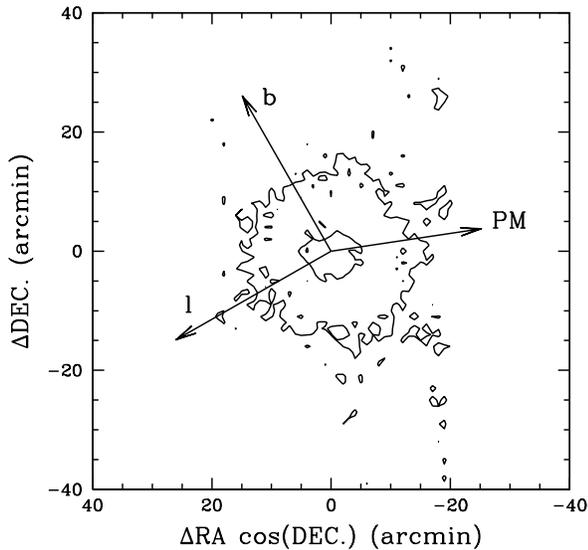, height=0.52\textwidth,width=0.52\textwidth}
 \vspace{-6mm}
 \caption{
Log scale density contour map for NGC 2506. 
The inner contour represents $f_{o}/2$ and the outer contour indicates 
$3\sigma$ level of background density fluctuation.
Each arrows represent the Galactic coordinate (l, b) and the proper 
motion vector (PM).
}
 \vspace{0mm}
 \label{density_contour}
\end{figure}

\subsection{Halo Structures}

To investigate the morphology and halo structure of the clusters, we plotted 
2 dimensional effective surface number density distributions of 
NGC 1245 and NGC 2506 in Fig. 9 and Fig. 10, respectively.
We plotted two contour levels. One corresponds to the effective surface 
number density of $f_{o}/2$ showing the representive morphology of the cluster
and the other corresponds to the $3\sigma$ of the background density ($f_{bg}$)
showing the outer morphology of the cluster 
which can be detected above the background fluctuations.
As can be seen in Fig. 9 and Fig. 10, NGC 1245 is more elongated
than NGC 2506. There seems to be no relation between the direction of the major 
axis of NGC 1245 and the direction of the cluster proper motion \citep{lok03}. 
We plotted the directions of cluster proper motion \citep{lok03} along with the directions of the Galactic latitude and longitude for comparison.
There are some features resembling tidal tails in the north-west of NGC 1245. 
The surface number densities of these features are comparable to the $3\sigma$ 
of the $f_{bg}$ which is $\sim1/8$ times the central surface density ($f_{o}$).

\section{SUMMARY AND CONCLUSIONS} 

We proposed a new method for determining the cluster membership probability 
using the spatial positions together with the positions in the CMD.
Owing to the ability to assign the membership probability for all the stars in 
the cluster field, we could trace the morphology of the clusters and 
derive the radial surface density profiles corrected for the field star 
contamination. 

The validity of the present method of 4-D membership probability is confirmed
by comparing our membership probabilities with those from proper motion
studies in the central region of NGC 2506 \citep{chi81}. However, our method
shows some bias to higher probability for stars in the central regions of
the cluster. This bias is caused by assigning high probabilities to stars in
the central regions of the cluster if they are located within the photometric
membership boundary in the CMDs.
We also tested the robustness of the segregation of non-member stars from 
the member stars by the Monte-Carlo simulations and found that the 4-D
method assigns membership probability quite convincingly. The probability of 
a star with $P>0.4$ is a real cluster member is about $90\%$.
However, for better understanding of the structure of open clusters,
we need kinematic data.

The two old open clusters NGC 1245 and NGC 2506 show an imprint of 
the dynamical evolution in their LFs and radial surface density profiles.
The LFs of the inner region of the clusters are dominated 
by the giant and upper main-sequence stars, whereas those of 
the outer region of the clusters are dominated by low-mass stars. 
This effect is more pronounced in NGC 2506.
Some fraction of low mass stars is expected to be evaporating
from the cluster to create the tidal halo surrounding the cluster. 
The over-dense regions revealed in the effective number density contour map 
of NGC 1245 suggest the presence of the tidal halo at $R > R_{cl}$ in the 
north-west of NGC 1245.
The detection of the over-dense regions at $R > R_{cl}$ 
was made possible due to our method of membership probability 
which took into account spatial positions together 
with the positions in the CMD, in addition to using 
wide and deep photometry of the two clusters (Paper I).

We found that the cluster core radius derived from bright stars is smaller
than that from the faint stars. 
The dependence of core radius on the luminosity of stars is 
a strong evidence of the mass segragation in NGC 1245 and NGC 2506.
Mass segregation and evaporation of the low mass stars are 
natural consequences of the dynamical evolution in relaxed clusters. 
Stars exchange their kinetic energy through encounters. 
Owing to the equipartition of the kinetic energy,
high-mass stars have relatively low velocity, whereas low mass stars may have
velocity that is sufficient to escape from the cluster.
If the evaporated stars are tidally disturbed by the Galactic disc or giant 
molecular clouds during their orbital motion, they are likely to be stretched
along the orbit \citep{chu10} to create tidal tails. A signature of tidal tail 
seems to be observed in NGC 1245 but the signal is too low to be confirmed.


\section*{Acknowledgments}
We would like to thank Prof. Kenneth Janes whose comments improved the
present paper significantly. 
This work was supported in part by the NRF research 2010-0023319
and supported in part by KASI(Korea Astronomy and Space Science Institute).

\clearpage

\end{document}